\documentclass[10pt]{article}
\usepackage[margin=1.5in]{geometry}

\usepackage[scaled]{helvet}
\usepackage[normalem]{ulem}

\usepackage{amsmath,bm}
\usepackage{graphicx,pdflscape}% Include figure files
\usepackage{color}
\usepackage[toc,page]{appendix}
\usepackage{simpler-wick}
\usepackage{subcaption}
\usepackage{float}
\usepackage{placeins}
\usepackage{hyperref}
\hypersetup{
    colorlinks,
    citecolor=red,
    filecolor=cyan,
    linkcolor=blue,
   urlcolor=magenta
 }

\usepackage{cleveref}

\usepackage{float}

\usepackage{cite}
\usepackage[normalem]{ulem}
\usepackage{xparse}
\ExplSyntaxOn
\NewDocumentCommand{\mref}{m}{\quinn_mref:n {#1}}
\seq_new:N \l_quinn_mref_seq
\cs_new:Npn \quinn_mref:n #1
 {
  \seq_set_split:Nnn \l_quinn_mref_seq { , } { #1 }
  \seq_pop_right:NN \l_quinn_mref_seq \l_tmpa_tl
  ( % print the left parenthesis
  \seq_map_inline:Nn \l_quinn_mref_seq
    { \ref{##1},\nobreakspace } % print the first references
  \exp_args:NV \ref \l_tmpa_tl % print the last or only one
  ) % print the right parenthesis
 }
\ExplSyntaxOff

\newcommand{\disp}[1]{Eq.~\mref{#1}}

\newcommand{\figdisp}[1]{Fig.~\mref{#1}}

\newcommand{\lessim} {\ {\raise-.5ex\hbox{$\buildrel<\over\sim$}}\ }

\newcommand{\gssim}{\ {\raise-.5ex\hbox{$\buildrel>\over\sim$}}\ }

\usepackage{color}

\newcommand{\si}{\sigma}
\newcommand{\sib}{\bar{\sigma}}
\newcommand{\tJ}{\ $t$-$J$ \ }
\newcommand{\nn}{\nonumber}

\renewcommand{\Re}{\mathrm{Re}}
\renewcommand{\Im}{\mathrm{Im}}
\newcommand{\Tr}{\mathrm{Tr}\,}

\renewcommand{\emph}{\textit}

\usepackage{color}

\newcommand{\beq}{\begin{eqnarray}}
\newcommand{\eeq}{\end{eqnarray}}
\newcommand{\barray}{\begin{eqnarray}}
\newcommand{\earray}{\end{eqnarray}}

\newcommand{\G}{{\cal G}}

\newcommand{\GH}{{\bf g}}
\newcommand{\GHI}{\GH^{-1}}

\newcommand{\overbar}{\bar}
\newcommand{\chem}{{\bm \mu}}

\renewcommand{\O}{{\cal O}}
\renewcommand{\Tilde}{\widetilde}
\newcommand{\wt}{\widetilde}

\newcommand{\half}{\frac{1}{2}}
\graphicspath{{UpdatedFigures2/}}

\begin{document}

\title{   Extremely Correlated Fermi Liquid theory for  $U=\infty$, $d=\infty$ Hubbard model to  ${\cal O}(\lambda^3)$ }

\author{ S Shears\footnote{sshears@ucsc.edu}, \, E Perepelitsky\footnote{edward.perepelitsky@gmail.com}, M Arciniaga\footnote{michael.arciniaga@gmail.com} \, and B S Shastry\footnote{sriram@physics.ucsc.edu}  \\
\em Physics Department, University of California, Santa Cruz, CA, 95064 }
\date{\today}

\maketitle

\abstract{We present the $\O(\lambda^3)$  results from the $\lambda$ expansion in the extremely correlated Fermi liquid theory applied to the infinite-dimensional \tJ model (with $J=0$), and compare the results with the earlier $\O(\lambda^2)$  results as well as the results from the dynamical mean field theory. We focus attention on the $T$ dependence of the  resistivity $\rho(T)$,  the Dyson self energy, and  the quasiparticle weight $Z$ at various densities. The comparison  shows that all  the methods  display quadratic in T resistivity followed by  a quasi-linear in T resistivity characterizing a strange metal, and  gives an estimate of the  different scales of  these variables relative to the exact results. }

\section{Introduction}

The \tJ model \disp{Eq-tJ} provides an important context  for understanding strongly correlated systems. 
It is closely related to the $U\gg t$ Gutzwiller-Hubbard-Kanamori \cite{Gutzwiller} model. It  is formally equivalent to the $U=\infty$  model to which we add superexchange interactions \disp{Eq-tJ}.  It can be obtained from a canonical transformation on the Hubbard model for large $U$, provided we throw out certain three center terms of ${\cal O}( t^2/U)$ \cite{Harris}.
In previous papers \cite{ECFL-0,ECFL-Main,ECFL-Pathintegral} we have developed the extremely correlated Fermi liquid (ECFL) theory to overcome the most difficult features of the model, namely the $U=\infty$ limit which eliminates a substantial fraction of states in the Hilbert space corresponding to double occupation of sites. The resulting  electrons are termed as Gutzwiller projected electrons, satisfying a non-canonical
algebra \disp{non-canonical}. As a result the Feynman diagram based perturbation theory fails here, and this motivated the development of the ECFL theory as described elsewhere.

We note that the importance of the physics of strong correlations  has motivated considerable activity in the theoretical community.
On the analytical side, the dynamical mean field theory (DMFT) \cite{ RMP,badmetal,HFL,Tremblay,Huangetal,Held,Vollhardt,Biermann } has matured into a reliable tool. It  uses  the numerical renormalization group of Wilson and Krishnamurthy\cite{Krishnamurthy,Krishnamurthy2} for a generalized Anderson impurity model, and  provides exact numerical results for the case of infinite dimension, which is the focus of the current work using ECFL.    

The ECFL theory is based on an expansion in a parameter $\lambda$ that is analogous to an expansion of magnetic system models such as the Heisenberg model,  for large spin, i.e. an expansion of relevant equations in powers of  $\frac{1}{2S}$. The main underlying observation  \cite{ECFL-Pathintegral,ECFL-Main,ECFL-0} is that the algebra of the Gutzwiller projected electrons is similar to that of the Lie algebra of spin operators, and hence allows the introduction of such a parameter that enables  a systematic expansion in powers of $\lambda$. The theory  
 has been developed so far using the $\O(\lambda^2)$ expressions for the  self energies in the problem, and applied in a variety of situations including $d=0$ i.e. the Anderson impurity model \cite{AndersonImpurity}, the $d=1$\tJ model \cite{ECFL-1d}, the $d=\infty, U=\infty$ Hubbard model \cite{ECFL-Dinfty-1,ECFL-Dinfty-2} and closest to experiments, the $d=2$\tJ model \cite{ECFL-2d-1,ECFL-2d-2,ECFL-2d-3}. At a formal level we have also established a systematic method for extending the expansion to high order terms, but in view of the additional  technical difficulties presented by them,  the effect of the higher order terms  have not yet been tested. 
This work reports the first results from the third order equations for the ECFL, applied to the case of the $d=\infty$ and $U=\infty$ Hubbard model. The results are compared with results from DMFT as well as with earlier 2nd order equations.

In Section.~(\ref{basic}) we summarize the basic aspects of the ECFL theory. We define the \tJ model and give an expression for the single electron Greens function of the Gutzwiller projected electrons, and the two self energies involved in the construction. We summarize the various approaches to the $\lambda$ expansion method, and explain the ideas behind the shift invariance of the equations, which are of great importance in the \tJ model. We then summarize the second and third order expressions for the self energies and cast these in a form that is convenient for computation.

In Section~(\ref{Section-second-sumrule}) we discuss the two sum-rules employed to fix the 
two Lagrange  multipliers in the problem. While one of them is the familiar particle number sum-rule, the other arises from the exact equations of motion of the Greens function, as derived in Appendix.~(\ref{Sumrule}). 

Section.~(\ref{Results}) presents the calculated results from the second and third order expansions. We focus attention on the $T$ dependence of the  resistivity $\rho(T)$,  the Dyson self energy, and  the quasiparticle weight $Z$ at various densities. We  compare these results between the successive approximations, the exact DMFT results and also the so called Tukey window  scheme used earlier by us. 

\section{Basic Theory \label{basic}}

%\section{Formulas}

%\subsection{2nd and 3rd Order}

\subsection{ECFL theory Formulas for $\G$ }

The general formalism underlying the theory of {\em extremely correlated Fermi liquids} (ECFL) has been discussed extensively in recent works\cite{ECFL-Main,ECFL-Pathintegral,ECFL-DMFT-1,ECFL-DMFT-2,ECFL-Lambda-Expansion}. Here  we record the  equations relevant to the present work, and point out the origin of the main equations in earlier works in  detail . We start here with the \tJ model
\beq
H= -\sum_{ij \si} t_{ij} \Tilde{C}_{i \si}^\dagger \Tilde{C}_{j \si} - \chem \sum_{i \si} \Tilde{C}_{i \si}^\dagger \Tilde{C}_{ i \si} + \frac{1}{2} \sum_{ij} J_{ij} \left( \vec{S}_i.\vec{S}_j - \frac{n_i n_j}{4} \right), \label{Eq-tJ}
\eeq
where $t_{ij}$ are the band hopping parameters whose  Fourier transform $\varepsilon_k$ is the band energy,   $n_i= \sum_{i \si} \Tilde{C}_{i \si}^\dagger \Tilde{C}_{ i \si}$  is   the number operator at site $i$, and   $\Tilde{C}^\dagger_{j \si},\Tilde{C}_{i \si}$ are Gutzwiller projected fermion operators\cite{ECFL-Main,ECFL-Pathintegral} satisfying non-canonical anticommutators
 \beq
 \{ \wt{C}_{i \si_i}, \wt{C}_{j \si_j}^\dagger \}=\delta_{ij} \left(\delta_{\si_i \si_j} - \si_i \si_j \wt{C}_{i \sib_i}^\dagger  \wt{C}_{i \sib_j} \right), \mbox{and  } \{ \wt{C}_{i \si_i}, \wt{C}_{j \si_j} \}=0.
 \label{non-canonical}
 \eeq

In ECFL theory, the Green's function for the \tJ model is given as the product of an auxiliary (canonical) Green's function $\textbf{g}$ and  a caparison function $\widetilde{\mu}$:
\begin{equation}
    \G(k,i\omega_m)=\textbf{g}(k,i\omega_m)\times\Tilde{\mu}(k,i\omega_m), \label{Eq-G}
\end{equation}
with the fermionic Matsubara frequency $\omega_m = k_BT (2 m+1) \pi$ and $k(=\vec{k})$ is the wave number.
These factors of $\G$ are expressed in terms of self energies $\Psi$ and $\Phi$ as
\beq
\widetilde{\mu}(k,i\omega_m)&=&1-\frac{n}{2} + \Psi(k,i\omega_m), \label{Eq-muhat}\\
\GH^{-1}(k,i\omega_m)&=& i \omega_m + \chem - (1-\frac{n}{2}) \varepsilon_k- \Phi(k,i\omega_m), \label{Eq-g}
\eeq
where $n$ is the number  of electrons of both spin per site
\beq
n= \langle n_i\rangle= \frac{2}{\beta N_s} \sum_{k, m} e^{i \omega_m 0^+} \G(k,i\omega_m) \label{Physical-Electron-Density}
\eeq
and $N_s$ is the number of sites in the lattice, and we set the lattice constant $a_0\to 1$.
Here $\Phi$ plays the role of  a Dyson-type self-energy for the canonical Green's function $\textbf{g}(k,i\omega_n)$ and $\Psi$ is a frequency dependent correction to $\widetilde{\mu}(k)$. 
These equations are valid in any dimension, and have been employed in different works in the special cases of $d=\infty$\cite{ECFL-DMFT-1,ECFL-DMFT-2,ECFL-Dinfty-1,ECFL-Dinfty-2,ECFL-Tukey}, $d=1$\cite{ECFL-1d} and $d=2$\cite{ECFL-2d-1,ECFL-2d-2,ECFL-2d-3}.

In this work we specialize to the $d=\infty$ case, which is convenient for the purpose of 
studying the systematics of the $\lambda$ expansion\cite{ECFL-Main,ECFL-Pathintegral,ECFL-Lambda-Expansion}. Here we set $J=0$ in \disp{Eq-tJ} and deal with what amounts to the $U=\infty$ Hubbard model.
In this limit  the self energies simplify\cite{ECFL-DMFT-1,ECFL-DMFT-2} to the following $k$ independent expressions
\begin{equation}
\Psi(k,i\omega_n)=\Psi(i\omega_n) \label{Eq-Psi}
\end{equation}
\begin{equation}
\Phi(k,i\omega_n)=\chi(i\omega_n)+\varepsilon_k \Psi(i\omega_n). \label{Eq-Phi}
\end{equation}
Here we observe  that the entire $\vec{k}$ dependence of $\Phi$ is  contained in the band energy $\varepsilon_k$. Therefore it follows that we can use $\varepsilon$ as a proxy for the wave vector. We  thus combine   \disp{Eq-muhat,Eq-g,{Eq-Phi}} to write
\begin{equation}
\GH^{-1}= i \omega_n + \chem - \Tilde{\mu}(i \omega_n) \varepsilon_k- \chi(k,i\omega_n). \label{Eq-g-2}
\end{equation}
This equation, together with \disp{Eq-muhat} and  \disp{Eq-G}  determines the physical Green's function $\G$. 
Combining them  we can formally write $\G$ in the standard Dyson form
\beq
\G(k,i\omega_n)&=& \frac{1}{\GH^{-1}_0(k,i\omega_n) - \Sigma_{D}(i\omega_n) } \label{G-Dyson}
\eeq 
where $\GH^{-1}_0(k,i\omega_n) = i \omega_n + \chem- \varepsilon_k$, and the manifestly $k$ independent Dyson self energy $\Sigma_D$ as
\beq
\Sigma_D(i \omega_n) = i \omega_n +\chem + \frac{\chi(i \omega_n) - i \omega_n - \chem}{1- \frac{n}{2} + \Psi(i \omega_n)} \label{Sigma-Dyson-1}.
\eeq
For later use we record the positive definite electron spectral function $\rho_{\G}(k,\omega)$ obtained by analytic continuation of $\G$ in \disp{G-Dyson}:
\beq
\rho_{\G}(k, \omega)= - \frac{1}{\pi} \Im \, \G(k,i\omega_n)\bigg|_{ i\omega_n \to \omega + i 0^+} .\label{spectral-G}
\eeq
In   experimental literature the spectral function  $\rho_{\G}(k,\omega)$ is denoted by $  A(k,\omega)$. Following  \disp{spectral-G} we define a spectral function obtained from $\GH$ following the same procedure
\beq
\rho_{\GH}(k, \omega)= - \frac{1}{\pi} \Im \, \GH(k,i\omega_n)\bigg|_{ i\omega_n \to \omega + i 0^+} .\label{spectral-g}
\eeq
Unlike $\rho_{\G}$, the variable  $\rho_{\GH}$ is a mathematical  object used in calculations that finally yield the physical spectral function $\rho_{G}$. We perform  calculations of the generalized  self energies $\chi,\Psi$  in a power series in $\lambda$ with coefficients that depend on $\GH$ (rather than $\G$), as described below. This expansion   determines the Dyson self energy $\Sigma_D$ through the rather complicated formula \disp{Sigma-Dyson-1}, which is in the form of a ratio of two expressions. This illustrates  the advantage of the ECFL formalism, which generates a highly non-trivial $\Sigma_D$, through relatively simple self energies $\chi,\Psi$ given below.

\subsection{ The $\lambda$-expansion:} A basic tool in the ECFL theory is an expansion of the fundamental, and in general intractable functional differential equations, in powers of a parameter $\lambda$ \cite{ECFL-0,ECFL-Main,ECFL-2}.  This parameter is a particular type of counter of the expansion, and  is set to unity after isolating  its different  powers in  the expansion of any physical quantity, such as the Greens function, or a self energy.  As explained in \cite{ECFL-Main} (see in particular Eqs.~(1,2,3,4,5)), the inspiration for the $\lambda$ expansion originally came from an  observation  in the case of the Hubbard model. Herein
 the entire set of  Feynman diagrams can be obtained by a similar expansion of exact functional differential equations in powers of the interaction constant   $U$. The strategy is then to find corresponding functional differential equations for the non-canonical Gutzwiller projected electrons of the \tJ model, and to invent a parameter that plays the role of $U$ in the Hubbard, albeit with a limited range. This program can be carried out systematically in three independent ways, as discussed next.

\begin{itemize}
\item[(A)]
\subitem$\bullet$ Term-by-term  iteration, i.e.  $\O(\lambda^{n})$ terms found by taking functional derivative of terms of $\O(\lambda^{n-1})$ \cite{ECFL-0,ECFL-Main,ECFL-Pathintegral}, i.e. 
 \subitem It can be introduced  as  a parameter in the exact Schwinger-Tomonaga functional differential equations determining the Green's functions \cite{ECFL-0,ECFL-Main}, followed by a systematic expansion of these equations\cite{ECFL-0,ECFL-Main,ECFL-Pathintegral,ECFL-Lambda-Expansion}.
The expansion itself can be done by taking successive functional derivatives of previous terms, as in \cite{ECFL-0,ECFL-Main,ECFL-Pathintegral}. 

\item[(B)] 
\subitem$\bullet$  Generalized diagrams of $\O(\lambda^n)$ \cite{ECFL-Lambda-Expansion}
\subitem Yet another method of expansion is through a diagrammatic expansion \cite{ECFL-Lambda-Expansion}, modeled after the Feynman graph representation of terms in the Schwinger-Tomonaga expansion. It brings in a new class of diagrams, outside the category described in Feynman diagrams, thanks to the non-canonical nature of the fermion algebra \disp{non-canonical}. The  paper \cite{ECFL-Lambda-Expansion} gives the  systematics of this procedure providing  rules extending  the Feynman diagram rules. With the help of the new set of rules,  one can write down expressions for terms to an arbitrary order $n$ without having to list terms of a lower order $n-1$.    This prior order listing is  mandatory   in the method [A], where we functionally differentiate terms of $\O(\lambda^{n-1})$ to generate terms of $\O(\lambda^n)$.

\item[(C)] \subitem$\bullet$ $\lambda$-fermions \disp{Lambda-canonical}, and their    equations of motion of to $\O(\lambda^n)$ \cite{ECFL-Pathintegral}.

\subitem Finally,  and most directly, we can introduce $\lambda$  through  a generalization of the anticommutation relations \disp{non-canonical}, by writing the anticommutators  \cite{ECFL-Pathintegral}
\beq
 \{ \wt{C}_{i \si_i}, \wt{C}_{j \si_j}^\dagger \}=\delta_{ij} \left(\delta_{\si_i \si_j} -  \, \lambda \, \si_i \si_j   \wt{C}_{i \sib_i}^\dagger  \wt{C}_{i \sib_j} \right), \label{Lambda-canonical}
 \eeq
where  $\lambda \in [0,1]$. 
These  anticommutators, together with  $ \{ \wt{C}_{i \si_i}, \wt{C}_{j \si_j} \}=0$ constitute  a Lie algebra that  defines  {\em  $\lambda$-fermions},  introduced in  (Ref.\cite{ECFL-Pathintegral} Sec.5).
At  $\lambda$=1 we recover  the Gutzwiller fermions
\disp{non-canonical}, while at $\lambda$=0 we recover  canonical  fermions. The introduction of these $\lambda$-fermions allows us to interpolate continuously between canonical fermions and Gutzwiller projected fermions.   The anticommutators are realized in terms of the canonical fermions using
 the correspondence \cite{ECFL-Pathintegral}
\beq
\wt{C}_{j \si_j}^\dagger\to {C}_{j \si_j}^\dagger (1- \lambda C_{j \sib_j}^\dagger C_{j \sib_j}), \; \wt{C}_{j \si_j} \to {C}_{j \si_j}, \; \mbox{ and  }   \wt{C}_{i \sib_i}^\dagger  \wt{C}_{i \sib_j} \to
 {C}_{i \sib_i}^\dagger  {C}_{i \sib_j}. \label{Lambda-fermion-rep}
\eeq
 The equations for the Green's functions for these $\lambda-$fermions can be similarly expanded systematically in powers of $\lambda$ leading to expressions for the twin self energies and other  objects to each order in $\lambda$

This procedure   has a  close parallel in  the familiar Kubo-Anderson spin-wave  expansion encountered in quantum magnets. In  the  version of that expansion, due to Freeman Dyson \cite{Dyson},   the usual angular momentum  Lie-algebra with spin-s: 
\beq
[S_i^\alpha,S_j^\beta]=i  \, \delta_{ij} \,  \varepsilon^{\alpha \beta \gamma} \; S_i^\gamma, \mbox{  and } \; \vec{S}_j.\vec{S}_j= s(s+1), \eeq
is realized using canonical bosons $b_i,b_i^\dagger$ their number operator $n_i=b_i^\dagger b_i$  with the correspondence
\beq
S_i^- = b_i, \; S_i^+= ( 2s) b_i^\dagger \left( 1- \frac{n_i}{ 2s} \right), \; \mbox{ and  } S_i^z= n_i-s \label{HP},
\eeq
together with a projection operator $P_D$  that ensures that the number of bosons per site is constrained to the finite number $n_i\leq 2s$. Proceeding in this way  Dyson and Maleev\cite{Dyson,Maleev}  showed that  a formal series in powers of $\frac{1}{2 S}$ is possible for physically relevant variables.  It is therefore clear that  this version of  spin wave expansion of quantum-magnets  is parallel to the $\lambda$-expansion of \disp{Lambda-canonical}  with the mapping $\lambda \leftrightarrow \frac{1}{2 S} $. Details and  references to  applications in quantum magnets using    this approach are     discussed in \cite{ECFL-Pathintegral}. 
\end{itemize}

 Finally it is worth mentioning that a  qualitative understanding of  this parameter $\lambda$ can be found in the simple context of a single site model. Here it  can be  seen explicitly that varying $\lambda$ in the range $\lambda\in[0,1]$ controls the fraction of double occupancy between its uncorrelated value and 0 (see \cite{ECFL-Main} Appendix-A).

\subsection{The Shift invariance and the second chemical potential $u_0$ \label{second-chempot} } 
At this stage we recall that the \tJ model has a simple invariance property
\beq
\chem\to \chem - \half u_0, \;\; \varepsilon_k \to \varepsilon_k - \half u_0. \label{Shift-Invariance}
\eeq
This property expresses the invariance of the band model, when the center of gravity of the band is shifted by an arbitrary constant $\half u_0$. We refer to $u_0$ as the second chemical potential of the problem,  requiring a second constraint in addition to the number sumrule \disp{Physical-Electron-Density}.   It becomes a strong  constraint in the ECFL theory, when we insist that \disp{Shift-Invariance}  should be preserved to each order in the $\lambda$ expansion. The freedom of choosing $u_0$ can be utilized to impose a subsidiary constraint on $\GH$, as discussed below.

 For imposing this invariance, we will accordingly shift both $\chem$ and $\varepsilon_k$ in \disp{Eq-g,Eq-Phi,Eq-g-2,Sigma-Dyson-1}. 
  With this change, and by incorporating the factors of $\lambda$ mentioned above, we   record the basic equations \disp{Eq-muhat,Eq-g,Eq-g-2} with a factor of $\lambda$ multiplying the relevant terms as well as the constant $u_0$ subtracted from $\chem$ as well as $\varepsilon_k$, as derived in \cite{ECFL-Tukey,FN-1} 
 \barray
 \widetilde{\mu}(k,i\omega_n)&=&1-\lambda \frac{n}{2} + \lambda \Psi(k,i\omega_n), \label{Eq-muhat-Lambda} \\
\GHI(k) &= &  i \omega_n + \chem- \frac{u_0}{2} - (\varepsilon_k - \frac{u_0}{2}) \widetilde{\mu}(i\omega_n)  - \lambda \,  \chi( i \omega_n) \label{Eq-g-3}
\earray
From \disp{Eq-muhat-Lambda}  it follows that when $\lambda\to0$, we get  $\Tilde{\mu}\to 1$ and hence  $\GH$ reduces to the non-interacting Green's function.
The task undertaken  in the next section is an expansion of this equation together with \disp{Eq-g-3} in powers of $\lambda$, giving  $\GHI$ and $\widetilde{\mu}$ to $\O(\lambda^3)$. Since $\Psi$ and $\chi$ have a prefactor of $\lambda$, their  expansion  to ($\O(\lambda)$) $\O(\lambda^2)$ generates an expansion to  ($\O(\lambda^{2})$) $\O(\lambda^{3})$ of $\wt{\mu}$ and $\GH^{-1}$. The $\O(\lambda^2)$ and  $\O(\lambda^3)$ expressions are taken from  \cite{ECFL-Tukey,FN-1}  and \cite{ECFL-Lambda-Expansion}.

%\subsection{ $\O(\lambda^2)$ and $\O(\lambda^3)$  Equations for the Greens functions}

\subsection{ The $\lambda$ expansion for the self energies }
We expand  the two  self energies  $\Psi,\chi$ (see \disp{{Eq-muhat},{Eq-muhat-Lambda}, {Eq-g-3}}) in powers of $\lambda$ as 
\beq
\Psi&=& \Psi_{[0]}+ \lambda \Psi_{[1]}+\lambda^2 \Psi_{[2]}+\ldots \\
\chi&=& \chi_{[0]}+ \lambda \chi_{[1]}+ \lambda^2 \chi_{[2]}+\ldots  \label{Lambda-Expansion-1} \eeq
which suffices to determine $\wt{\mu}$ and $\GH^{-1}$ to $\O(\lambda^3)$.
We first  record the lowest order terms \cite{ECFL-Lambda-Expansion}
\beq\Psi_{[0]}=0,\eeq
\beq \chi_{[0]}= - \sum_p \GH(p) e^{i \omega_p 0^+} (\varepsilon_p - \frac{u_0}{2})= \frac{u_0 n_g}{4}-  \sum_p \GH(p) \varepsilon_p e^{i \omega_p 0^+},  \label{Zeroeth-Order}   \eeq
 where  $\sum_p \equiv \frac{k_BT}{N_s} \sum_{\vec{p} \omega_p}$, and $N_s$ is the number of lattice sites.  We defined $n_g$ using
\beq
  n_g= 2 \sum_p \GH(p) e^{i \omega_p 0^+}.  \label{ng}
\eeq
For brevity  the  factor $e^{i \omega_p 0^+}$ is omitted in the following, whenever we sum over a single $\GH$.
Here  $n_g$ is a formal construct and should not be confused with  the number density of physical electrons $n$, the latter is given in terms of $\G$ in \disp{Physical-Electron-Density}.   In practice $n_g$   turns out to be  quite close to $n$ at low $T$. %\figdisp{ng-nG}. 

Incorporating the terms in \disp{Zeroeth-Order}, we write
 \barray
 \widetilde{\mu}(k,i\omega_n)&=&1-\lambda \frac{n}{2} + \lambda^2 \Psi_{[1]}(k,i\omega_n)+ \lambda^3 \Psi_{[2]}(k,i\omega_n) +\O(\lambda^4), \label{Eq-muhat-Lambda-1}  \\
\GHI(k) &= &  i \omega_n + \chem' - (\varepsilon_k - \frac{u_0}{2}) \widetilde{\mu}(i\omega_n)  - \lambda^2 \,  \chi_{[1]}( i \omega_n) - \lambda^3 \,  \chi_{[2]}( i \omega_n)+\O(\lambda^4) \nn \\\label{Eq-g-4}
\earray
where 
\beq
\chem'&=&\chem - \frac{u_0}{2}- \lambda \chi_{[0]}\nn \\
&=&\chem -\frac{u_0}{2} - \lambda \left(\frac{u_0 n_g}{4} - \sum_p \GH(p) \varepsilon_p \right). \label{Eq-28}
\eeq
In the sum-rule \disp{Second-sumrule} we require the true $\chem$ obtained from $\chem'$. For  this purpose we use the  expression 
\beq
\chem=\chem'+ \frac{u_0}{2}\left( 1+ \frac{n}{2}\right) -  \sum_p \GH(p) \varepsilon_p \label{chemical}
\eeq
obtained 
after setting $\lambda=1$.   We make an extra technical assumption of replacing $n_g$ in \disp{Eq-28} with $n$,  the   particle density in order to accelerate convergence.  The resulting  spectral functions  using  $n$  are very close to those  using  $n_g$ whenever both methods converge.
In order to completely define the scheme \disp{{Eq-muhat-Lambda-1},{Eq-g-4}} we need formal expressions for  $\Psi_{[j]}(k)$ and $\chi_{[j]}(k)$ with $j=1,2$. They are given as functions of $k$ below, and analyzed later to show that these are independent of the wave vector $\vec{k}$ and functions {\em only} of the Matsubara frequency $ \omega_k = \frac{\pi}{\beta}(2k+1)$.

\subsubsection{Second Order}

 The second order  $\lambda$ expansion  \{see Eqs.~(10,11) in \cite{ECFL-Tukey}\} gives us the following two self-energy parts:
\beq \Psi_{[1]}(k)= - \sum_{pq} (\varepsilon_p+\varepsilon_q - u_0)\GH(p) \GH(q) \GH(p+q-k) \;\; \label{Psi-1} \eeq
\beq \chi_{[1]}(k)= - \sum_{pq} (\varepsilon_{p+q-k}- \frac{u_0}{2})) (\varepsilon_p+\varepsilon_q - u_0)\GH(p) \GH(q) \GH(p+q-k) \;\; \label{chi-1} \eeq

\subsubsection{Third Order}

 The third order  $\lambda$ expansion  \{see Eqs.~(65:b-g) with $J$=$0$ in \cite{ECFL-Lambda-Expansion}\} gives us $\Psi_2$ as
\barray
 \Psi_{[2]}(k)&=&  \ -4  \sum_{pql}\GH(p)\GH(l)\GH(q)\GH(k+q-p)\GH(k+q-l)(\varepsilon_p-\frac{u_0}{2})(\varepsilon_l-\frac{u_0}{2})\nn\\
&&  \ -\sum_{pql}\GH(p)\GH(l)\GH(q)\GH(k+q-p)\GH(q+l-p)(\varepsilon_p-\frac{u_0}{2})(\varepsilon_l-\frac{u_0}{2})\nn\\
&&  \ - \sum_{pql}\GH(p)\GH(l)\GH(q)\GH(k+q-p)\GH(p+l-q)(\varepsilon_p-\frac{u_0}{2})(\varepsilon_q-\frac{u_0}{2})\nn\\
&&  \ -  \sum_{pql}\GH(p)\GH(l)\GH(q)\GH(k+q-p)\GH(k+l-p)(\varepsilon_p-\frac{u_0}{2})(\varepsilon_q-\frac{u_0}{2})\nn\\
&&  \ -\sum_{pql}\GH(p)\GH(l)\GH(q)\GH(k+q-p)\GH(l+p-k)(\varepsilon_p-\frac{u_0}{2})(\varepsilon_l-\frac{u_0}{2})\nn\\
&& +  \  \frac{n}{2}\sum_{pq}\GH(p)\GH(q)\GH(k+q-p)(\varepsilon_p-\frac{u_0}{2}) \label{Psi-2}
\earray
while $\chi_{[2]}(k)$ is given   \{see Eqs.~(66:b-g)  with $J$=$0$ in \cite{ECFL-Lambda-Expansion}\} by the sum of the following terms:

\barray
\chi_{[2]}(k)&=&  \ -  \sum_{pql}\GH(p)\GH(l)\GH(q)\GH(k+q-p)\GH(l+q-p)(\varepsilon_p-\frac{u_0}{2})(\varepsilon_l-\frac{u_0}{2})(\varepsilon_{l+q-p}-\frac{u_0}{2})\nn\\
&&  \ -4 \sum_{pql}\GH(p)\GH(l)\GH(q)\GH(k+q-p)\GH(k+q-l)(\varepsilon_p-\frac{u_0}{2})(\varepsilon_l-\frac{u_0}{2})(\varepsilon_q-\frac{u_0}{2})\nn\\
&&  \ - \sum_{pql}\GH(p)\GH(l)\GH(q)\GH(k+q-p)\GH(k+l-p)(\varepsilon_p-\frac{u_0}{2})(\varepsilon_l-\frac{u_0}{2})(\varepsilon_q-\frac{u_0}{2})\nn\\
&&  \ -  \sum_{pql}\GH(p)\GH(l)\GH(q)\GH(k+q-p)\GH(p+l-q)(\varepsilon_p-\frac{u_0}{2})(\varepsilon_q-\frac{u_0}{2})(\varepsilon_l-\frac{u_0}{2})\nn\\
&&  \ -  \sum_{pql}\GH(p)\GH(l)\GH(q)\GH(k+q-p)\GH(k+l-p)(\varepsilon_p-\frac{u_0}{2})(\varepsilon_l-\frac{u_0}{2})(\varepsilon_{k+l-p}-\frac{u_0}{2})  \nn \\ \label{Chi-2}
\earray
By Fourier transforming in space, i.e. by going to real space, it is readily seen that the dependence on $\vec{k}$ drops off and hence both $\Psi$ and $\chi$ are only functions of the frequency $i\omega_n$. This Fourier transformation is facilitated by observing that most  factors of $\GH$ are   accompanied by a corresponding factor of $(\epsilon_p-\frac{u_0}{2})$ of the same momentum $p$.  

\subsection{Further Simplification of formulae}
These  formulae \disp{Psi-1,chi-1,Psi-2,Chi-2} can then be expressed more simply in terms of the following objects:
\beq
\GH_{loc,m}(i\omega_k)&\equiv& \frac{1}{N_s} \sum_{\vec{k}} \left( \varepsilon_k \right)^m \; \GH(k,i\omega_k)\\
\GH_0(i\omega_k) &\equiv& \GH_{loc,0}(i\omega_k); \label{Eq-35} \\
 \GH_1(i\omega_k)&\equiv& \GH_{loc,1}(i\omega_k) - \frac{u_0}{2}\;\GH_{loc,0}(i\omega_k) 
\eeq
and the bilinear objects
\beq \gamma_{m,n}(i\Omega_k)\equiv \frac{1}{\beta} \sum_{\omega_p}\GH_m(i\omega_p)\GH_n(i\Omega_k-i\omega_p), \\
\zeta_{m,n}(i\Omega_k)\equiv \frac{1}{\beta} \sum_{\omega_p}\GH_m(i\omega_p)\GH_n(i\Omega_k+i\omega_p).
\eeq
We caution the reader that in this paper the object $\GH_0$ is defined in \disp{Eq-35}. It is {\em different} from a non-interacting greens function, as sometimes denoted in literature.
We note the symmetries $\gamma_{m,n}(i\Omega_k) = \gamma_{n,m}(i\Omega_k)$ and $\zeta_{m,n}(i\Omega_k) = \zeta_{n,m}(-i\Omega_k)$.  In the diagonal case of $m=n$, these definitions and symmtries reduce to the standard identities  for the ``bubble" diagram. The formulae for $\psi$ and $\chi$ then become
\barray
 \Psi_{[1]}(i\omega_k) &=& - 2 \frac{1}{\beta} \sum_{\omega_q} \gamma_{1,0}(i\omega_k + i\omega_q)\GH_0(i\omega_q) \label{Psi-First-Order-Final}\\
\Psi_{[2]}(i\omega_k) &=&  - 4 \frac{1}{\beta} \sum_{\omega_q} \gamma^2_{1,0}(i\omega_k + i\omega_q)\GH_0(i\omega_q) 
- \frac{1}{\beta} \sum_{\omega_q} \zeta^2_{0,1}(i\omega_k - i\omega_q)\GH_0(i\omega_q) \nn\\
&& - \frac{1}{\beta} \sum_{\omega_q} \zeta_{1,1}(i\omega_k - i\omega_q)\zeta_{0,0}(i\omega_k - i\omega_q)\GH_0(i\omega_q) \nn\\
&& - \frac{1}{\beta} \sum_{\omega_q} \zeta_{1,0}(i\omega_k - i\omega_q)\zeta_{0,0}(i\omega_k - i\omega_q)\GH_1(i\omega_q) \nn\\
&&- \frac{1}{\beta} \sum_{\omega_q} \zeta_{0,0}(i\omega_k - i\omega_q)\zeta_{0,1}(i\omega_k - i\omega_q)\GH_1(i\omega_q)\nn\\
&& +\frac{1}{\beta} \frac{n}{2} \sum_{\omega_q} \zeta_{0,1}(i\omega_k - i\omega_q)\GH_0(i\omega_q)  \label{Psi-Second-Order-Final}
\earray
and 
\barray
 %\chi_{[0]}(i\omega_k) &=& \sum_{i\omega_p}\GH_1(i\omega_p) \nn\\
 \chi_{[1]}(i\omega_k) &=& - 2\frac{1}{\beta} \sum_{\omega_q} \gamma_{1,0}(i\omega_k + i\omega_q)\GH_1(i\omega_q) \label{Chi-First-Order-Final} \\
\chi_{[2]}(i\omega_k) &=&  - \frac{1}{\beta}\sum_{\omega_q} \zeta_{0,1}(i\omega_k - i\omega_q)\zeta_{1,1}(i\omega_k - i\omega_q)\GH_0(i\omega_q)\nn\\
&&- 4 \frac{1}{\beta}\sum_{\omega_q} \gamma^2_{1,0}(i\omega_k + i\omega_q)\GH_1(i\omega_q) 
- \frac{1}{\beta}\sum_{\omega_q} \zeta^2_{1,0}(i\omega_k - i\omega_q)\GH_1(i\omega_q) \nn\\
&& -\frac{1}{\beta} \sum_{\omega_q} \zeta_{1,1}(i\omega_k - i\omega_q)\zeta_{1,0}(i\omega_k - i\omega_q)\GH_0(i\omega_q) \nn\\
&& -\frac{1}{\beta} \sum_{\omega_q} \zeta_{1,1}(i\omega_k - i\omega_q)\zeta_{0,0}(i\omega_k - i\omega_q)\GH_1(i\omega_q) \label{Chi-Second-Order-Final} 
\earray
Substituting the expressions  \disp{Psi-First-Order-Final,Chi-First-Order-Final,Psi-Second-Order-Final,Chi-Second-Order-Final}   in \disp{Eq-muhat-Lambda-1,Eq-g-4} and setting $\lambda=1$, we obtain the basic equations to third order in $\lambda$. To get the corresponding second order equations we simply drop the third order terms \disp{Psi-Second-Order-Final,Chi-Second-Order-Final}.

\section{ Fixing ${ \mu}$ and $u_0$   \label{Section-second-sumrule}  }
The numerical evaluation  of \disp{Eq-muhat-Lambda-1,Eq-g-4} begins   after setting $\lambda=1$ in these equations.   We need two constraints  to  
 determine the two  parameters $\chem$ (or $\chem'$) and $u_0$ (see Sec.~\ref{second-chempot}).
The  sum-rule 
\beq
n_{\G}=  2 \sum_p\G(p) e^{i \omega_p 0^+} = n, \label{First-sumrule}
\eeq
 which is equivalent to  $=2\sum_k \int d\omega \rho_{\G}(k,\omega) f(\omega) $ \disp{Physical-Electron-Density} fixes the total electron density. The factor of $2$ arises from spin summation. For the second sum-rule  there are two alternatives as noted next.

\begin{itemize}
\item 
In our earlier work   \cite{ECFL-0,ECFL-Main}  we  imposed another sum-rule
\beq
n_g\; = 2 \sum_p\GH(p) e^{i \omega_p 0^+}=n  \label{Second-ng-Sumrule}.
\eeq
At low $T$ this sum-rule can be argued for using  the Luttinger-Ward theorem (see Eq.~(16) in \cite{ECFL-0} ) at low $T$, and  in the absence of  alternatives at all $T$.  For electrons at densities $0.7\leq n\lessim 1$,  by enforcing this sum-rule, the spectral functions    generate (low amplitude) tails spread over  very high energies. These tails  are unexpected on physical grounds and are thus unwanted. In order to curtail these tails, a Tukey-type energy window was introduced in \cite{ECFL-Tukey} (see Eq.~(33,34)). This  window  cuts off  the  high energy tails,   and we  then renormalize  the spectral weight inside the window to satisfy the unitary sum-rule $\int \rho_\GH(k,\omega) d \omega= 1$ at each $k$ \cite{comment-1}. This procedure leads to compact spectral functions that seem physically reasonable. They  compare  reasonably with exact results from DMFT at low T and low $\omega$,  as shown in \cite{ECFL-Tukey} and later in  \cite{ECFL-Dinfty-1,ECFL-Dinfty-2}. We shall refer to  spectral functions obtained using  \disp{Second-ng-Sumrule}, and the above energy windows, as  the Tukey window scheme results. These  are  displayed  below in \figdisp{Figure1,Figure2} at relevant densities. 

\item
In this work we study an alternate method where we impose  a different   sum-rule from the earlier ones.    The sum-rule used  is an   exact relation that the spectral function must satisfy, given the Hamiltonian of the system and the standard commutation relations. 
The details of its derivation can be found in Appendix (\ref{Sumrule}).
 In the case of infinite dimensions where the exchange energy $J=0$ we find the exact sum-rule 
\beq
\sum_k \int d\omega \rho_{\G}(k,\omega) f(\omega) (\omega+ \chem- \varepsilon_k) =0, \label{Second-sumrule}
\eeq
where $f(\omega)=\bigg(1+e^{\beta \omega}\bigg)^{-1}$ is the Fermi function.
For the record we also note the sum-rule for the  model on the 2-d square lattice with a finite $J$. Here the exact expression for the exchange energy is not known, and we quote the result from
 a Hartree approximation:
\beq
\sum_k \int d\omega \rho_{\G}(k,\omega) f(\omega) (\omega+ \chem- \varepsilon_k) =- \frac{J n^2}{2},  \label{Second-sumrule-2D}
\eeq
This sum rule has been used throughout this paper for our second and third order code, and the results are compared with earlier ones where the Tukey window cutoff was  used.

\item   We found in several tests that solutions found any two of these sum-rules already seems to satisfy  the third one reasonably well, but not exactly so.  While using \disp{Second-ng-Sumrule} is attractive at low-T since it captures the Luttinger-Ward Fermi surface exactly, it does create long tails extending to high energies requiring further cutoff schemes such as the Tukey window discussed in  \cite{ECFL-Tukey}.  
In order to explore other possibilities, we avoid using this sum-rule.
In the present work  only  the  \disp{First-sumrule} and  \disp{Second-sumrule} are used. See Appendix (\ref{Code}) for further details.

\end{itemize}

\section{ Results and Discussion \label{Results}}

Let us first summarize the steps followed in this calculation of  
the solution of the ECFL equations to  $\O(\lambda^3)$. The $\O(\lambda^2)$ calculation follows by neglecting the third order terms. The task is to    solve
\disp{Eq-muhat-Lambda-1,Eq-g-4} for $\GH,\widetilde{\mu}$ after setting $\lambda=1$, with $\Psi_{[j]},\chi_{[j]}$ with $j=1,2$  given by \disp{Psi-First-Order-Final,Psi-Second-Order-Final}. Here   $\GH$, $\Psi$ and $ \; \widetilde{\mu}$ are calculated  from \disp{Eq-muhat-Lambda-1,Eq-g-4} 
in terms  of $\Psi,\chi$, which are given in terms of $\GH$,  $\Psi$   \disp{Chi-First-Order-Final,Chi-Second-Order-Final}- thus forming a self consistent non-linear set of equations for these functions. The external parameters needed for this calculation are the density $n$ and the temperature $T$, while the internal parameters are $\chem$ and $u_0$.    As discussed in Sec.~(\ref{Section-second-sumrule}), in the present work these  internal parameters are determined using \disp{First-sumrule,Second-sumrule}. Eqs.~(\ref{Chi-First-Order-Final},\ref{Chi-Second-Order-Final}) are expressible as convolutions of suitable functions and can be  efficiently  evaluated using fast Fourier transforms. 

The calculations in $d=\infty$ are performed using the popular Bethe lattice semicircular density of states
\beq
\rho_{DOS}(\varepsilon) = \frac{2}{\pi D^2} \sqrt{D^2-\varepsilon^2}
\eeq 
so that $D$ is the half band width usually estimated as $D\sim\O(1)$ eV, i.e. $D\sim10^4$ K.
The  calculations presented here are at  temperatures $T\leq 0.1 D$, and are the first ones using the new $u_0$ sum rule \disp{Second-sumrule}.

\begin{figure}%[htp]
\centering
\includegraphics[width=.7\textwidth]{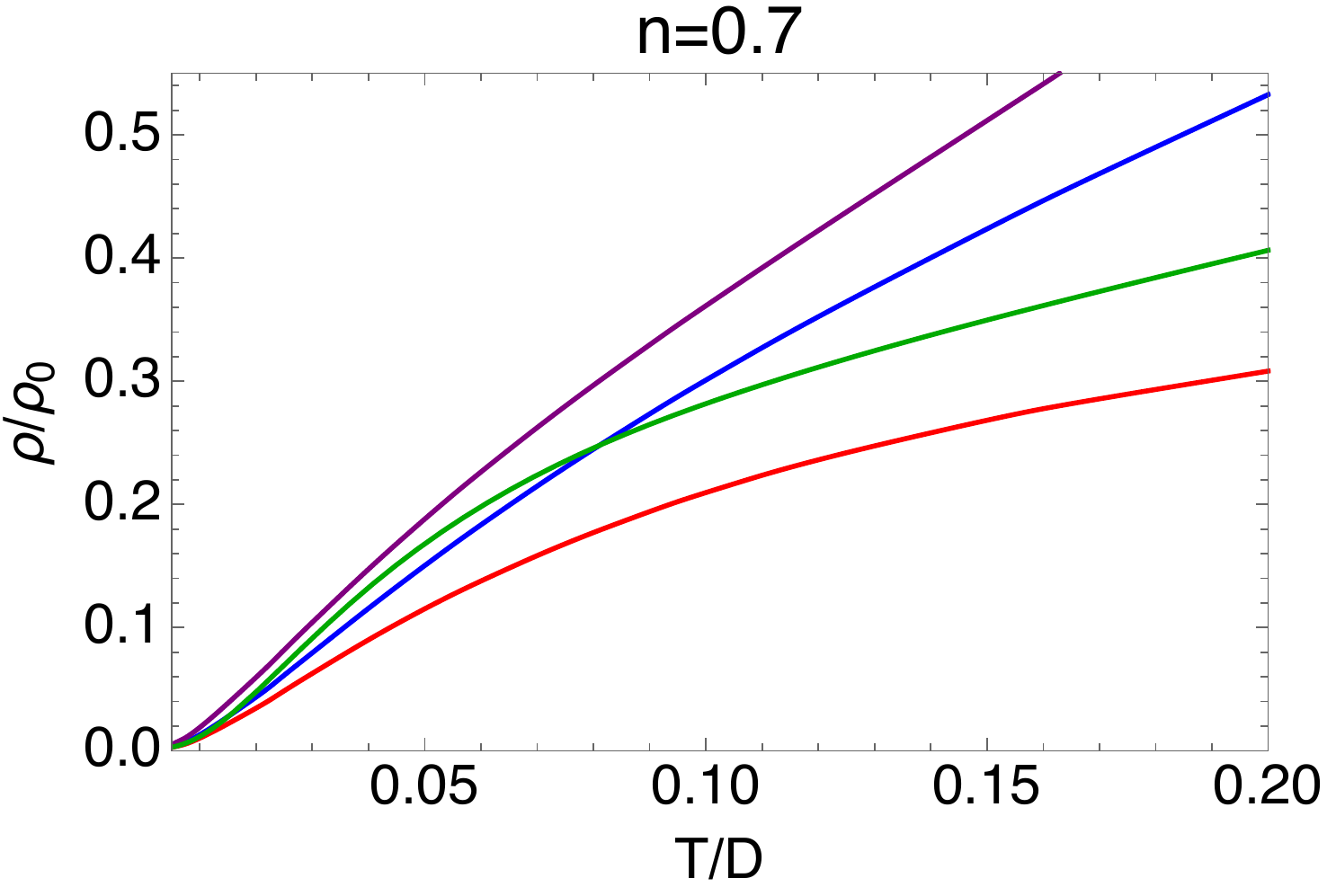}\hfill
\caption{\footnotesize Resistivity plots for density 0.7 to T=0.2. Here the resistivity  $\rho_0$ is  defined \cite{ECFL-Tukey} (see Eq.~(40)) as the inverse of the characteristic conductivity $\sigma_0= e^2 \hbar \Phi(0)/D$, with $\Phi(0)= \frac{1}{a_0^3}\rho_{DOS}(0) \langle (v_k^x)^2\rangle_{\varepsilon_k=0}$ and $a_0$ the lattice constant.
 The red plots are second order with the $u_0$ sum rule, the blue are third order with the $u_0$ sum rule, the green are the DMFT results (using an extrapolation from higher density results) and the purple are second order without the $u_0$ sum rule using the Tukey window scheme. In this and other  subsequent plots the DMFT results were kindly provided by Professor Rok \v{Z}{i}tko. }
\label{Figure1}
\end{figure}

In \figdisp{Figure1} we display  the resistivity at $n=0.7$ for $0\leq T\leq 0.2 D$ from the second (red) and third order (blue) calculations using \disp{Second-sumrule}, and compare with the exact DMFT results (green) at $U=\infty$ and $d=\infty$ for these parameters. We also display the results (purple) from the second order Tukey window scheme (i.e. using the \disp{Second-ng-Sumrule} together with the Tukey window). These are seen to be close to the exact DMFT result for $T\lessim 0.05 D$, and somewhat overshoot  the other estimates as we raise $T$. Both the second and third order results obtained using \disp{Second-sumrule} (red and blue curves), show a quadratic in T behavior (i.e. $ \rho \propto T^2$)  for $T\lessim 0.02 D$. This is  similar to the behavior of the  exact DMFT curve (green). At higher $T$ ( say $T \lessim 0.05 D$) both curves display a quasi-linear regime $\rho \propto T$, which is sometimes referred to as the ``strange-metal'' regime.  At even higher $T$, these two curves separate out. In general the third oder curve (blue) is closest to the exact DMFT result (green) over the entire $T$ regime.  The DMFT results, however, display a bend and subsequent second quasi-linear regime with a different slope and zero-intercept relative to the first, as the temperature increases above $T\gssim 0.10 D$. While present to some extent in all three ECFL calculations, it is most pronounced in the second order curve (red).
\begin{figure}%[htp]
\centering
\includegraphics[width=.5\textwidth]{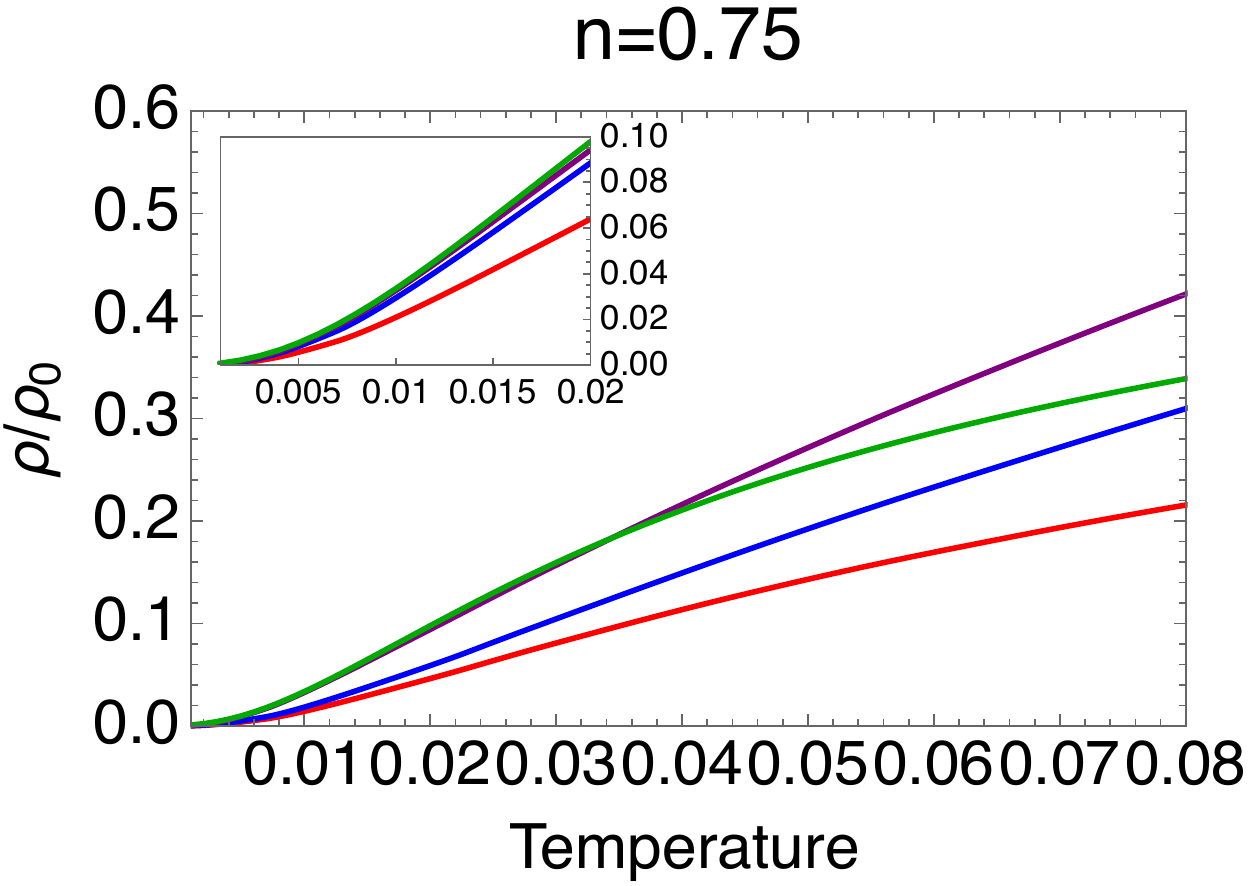}\hfill
\includegraphics[width=.5\textwidth]{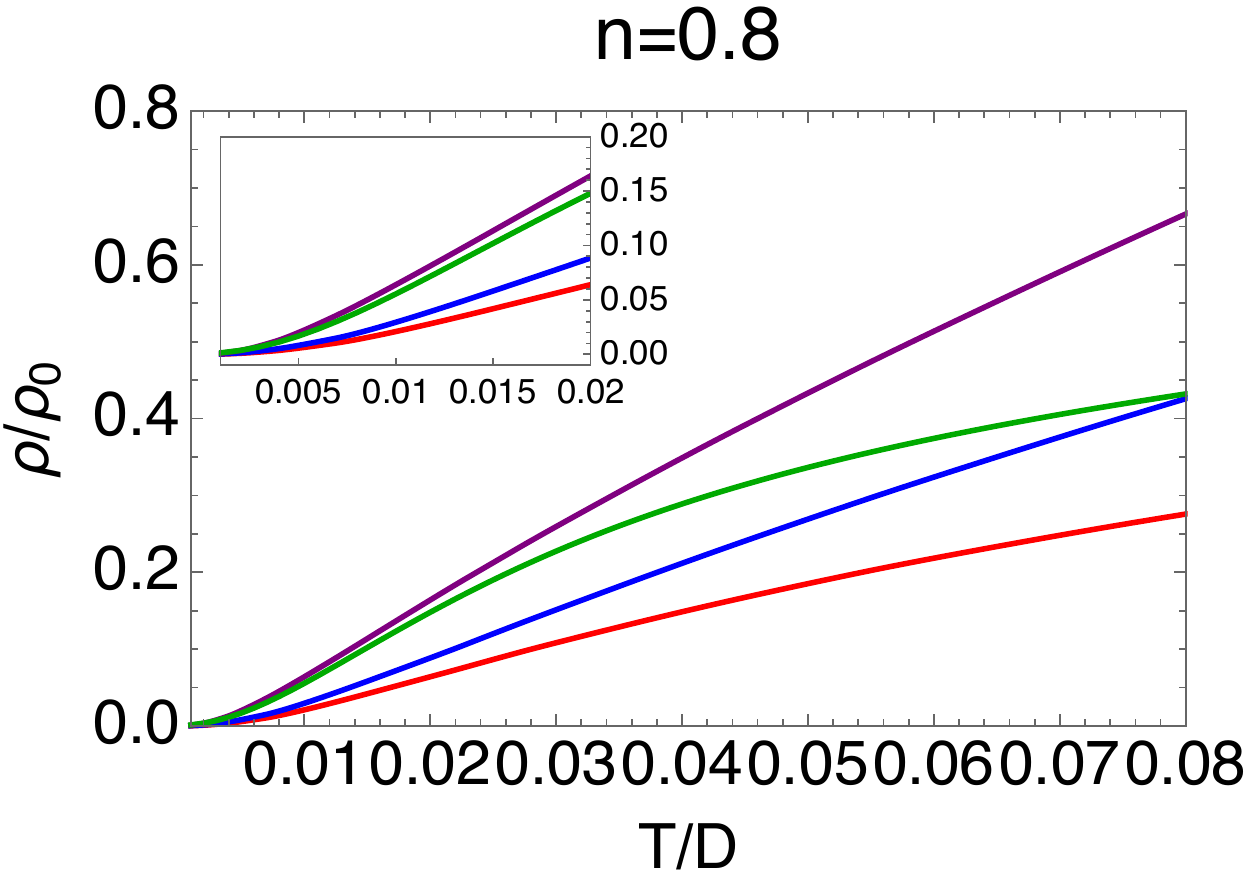}\hfill
\includegraphics[width=.5\textwidth]{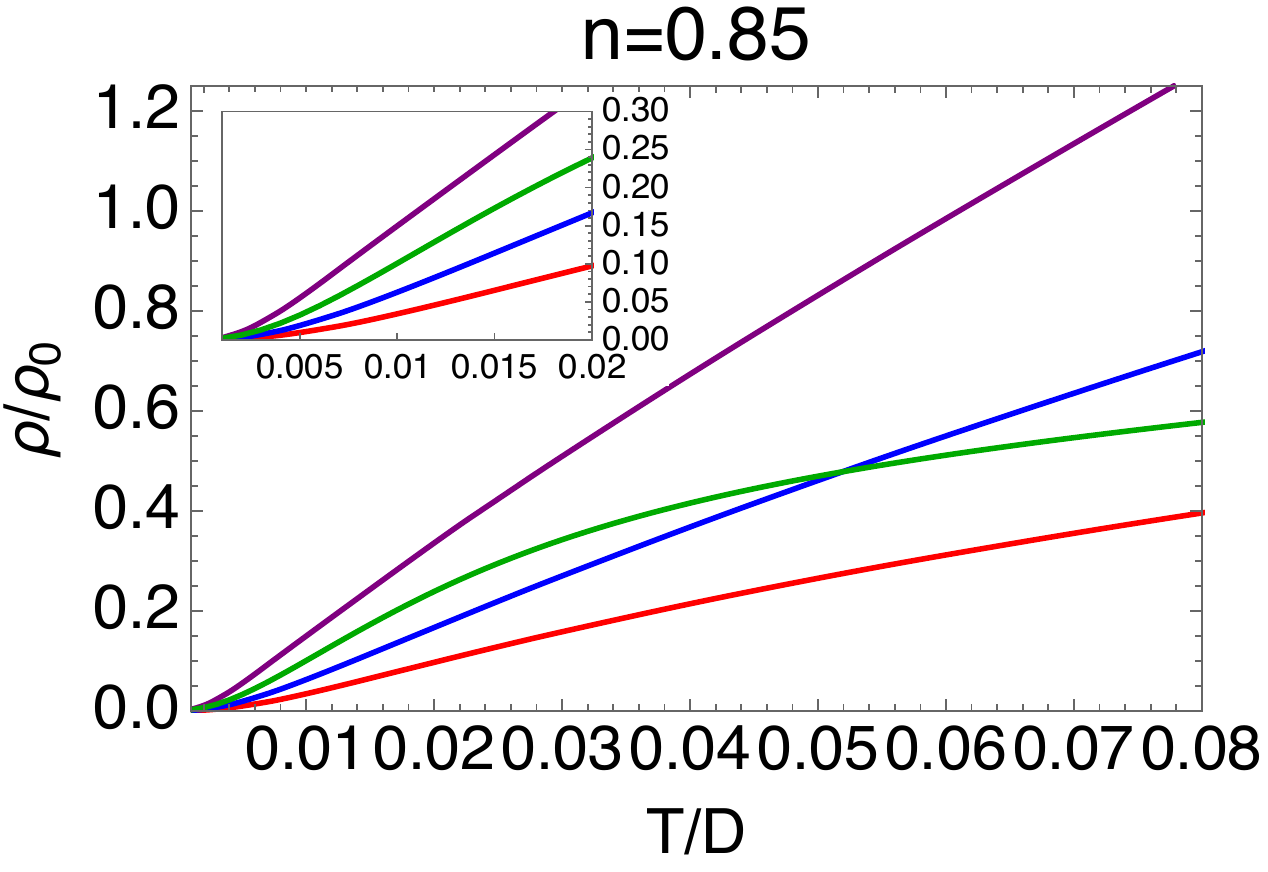}
\caption{\footnotesize Resistivity plots for densities 0.75, 0.8 and 0.85.  The resistivity $\rho_0$ is defined in the caption of \figdisp{Figure1}. The red plots are second order with the $u_0$ sum rule, the blue are third order with the $u_0$ sum rule, the green are the DMFT results and the purple are second order without the $u_0$ sum rule using the Tukey window scheme. The insets show the resistivity on a smaller temperature scale.}
\label{Figure2}
\end{figure}

In \figdisp{Figure2} we compare the resistivities obtained from the second order scheme (red), the third order scheme (blue), the Tukey window scheme (purple), and the exact DMFT results (green) at higher electron densities $n$, i.e. lesser hole doping $\delta=1-n$.  The insets show the comparison at very low $T\lessim 0.02 D$ and the main figures present a larger regime $0\leq T \leq 0.08 D$. In going from second to third order, we see that the resistivities are closer to the DMFT results  at all densities. The Tukey window scheme on the other hand, is quite close to DMFT for $n<0.85$, while for $n=0.85$ it becomes an overcorrection.

\begin{figure}[!ht]
\centering
\includegraphics[width=.33\textwidth]{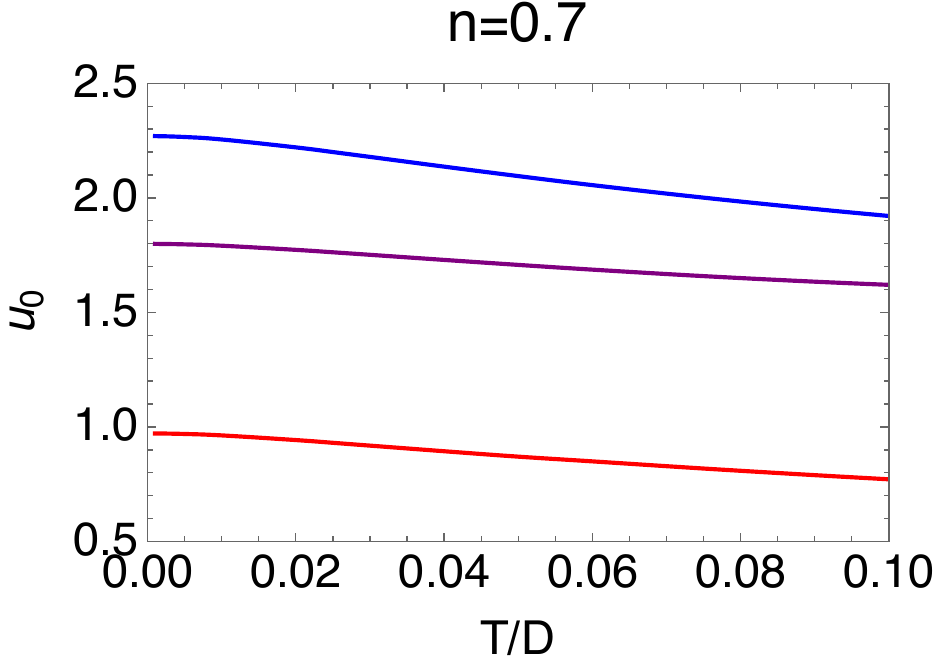}\hfill
\includegraphics[width=.33\textwidth]{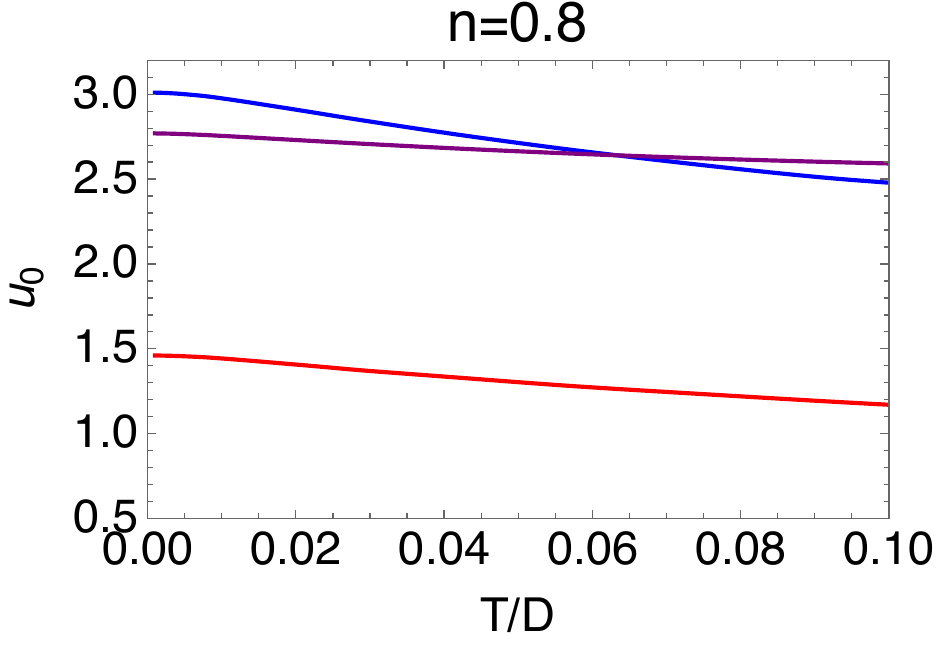}\hfill
\includegraphics[width=.33\textwidth]{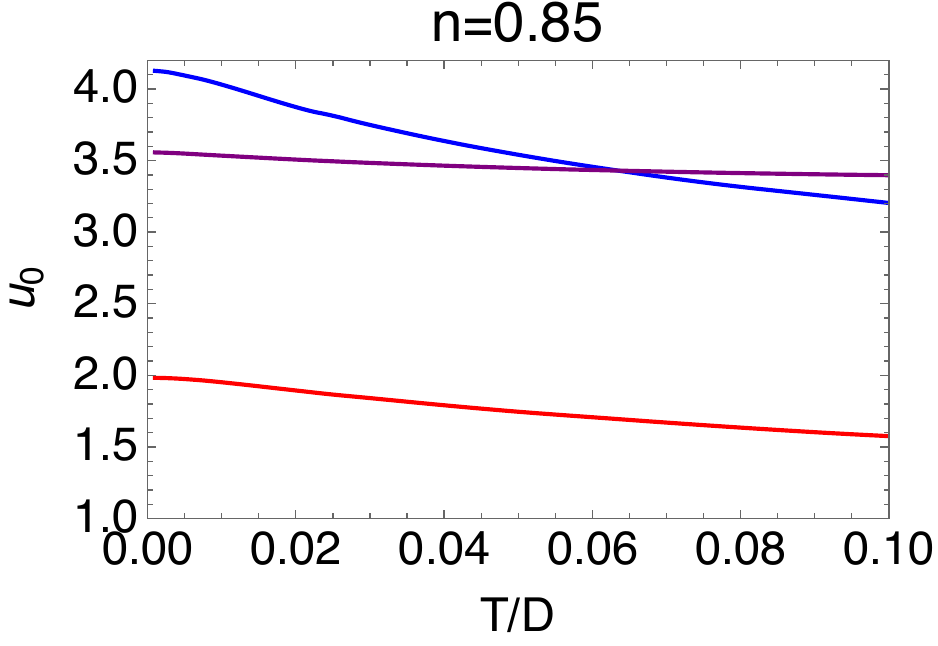}
\caption{\footnotesize  The second chemical potential $u_0$ (see Sec.~\ref{second-chempot})  for densities 0.7, 0.8 and 0.85. The red plots are second order with the $u_0$ sum rule, the blue are third order with the $u_0$ sum rule and the purple are second order without the $u_0$ sum rule using the Tukey window scheme.}
\label{Figure3}
\end{figure}

In \figdisp{Figure3} we display  the second chemical potential $u_0$ in the second and third order results and compare those with the Tukey window scheme results. Results are shown only up to $n=0.85$ since upon going  past this limit,    the third order $u_0$ grows beyond $\sim 4 D$ rendering the convergence of the scheme as somewhat  unstable.  

%\FloatBarrier

\begin{figure}[htp]
\centering
\includegraphics[width=.5\textwidth]{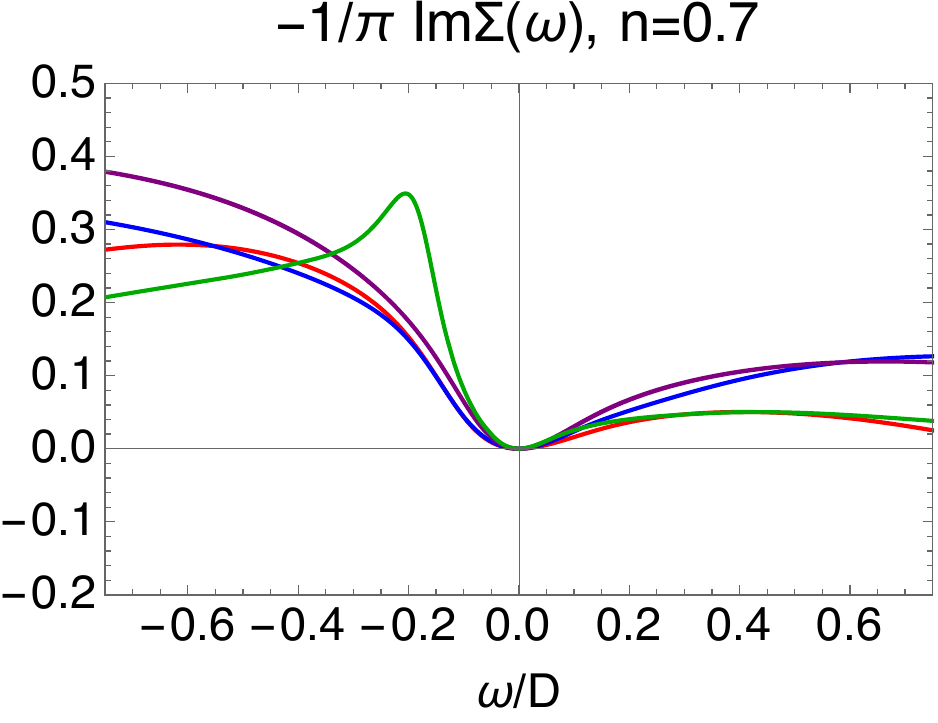}\hfill
\includegraphics[width=.5\textwidth]{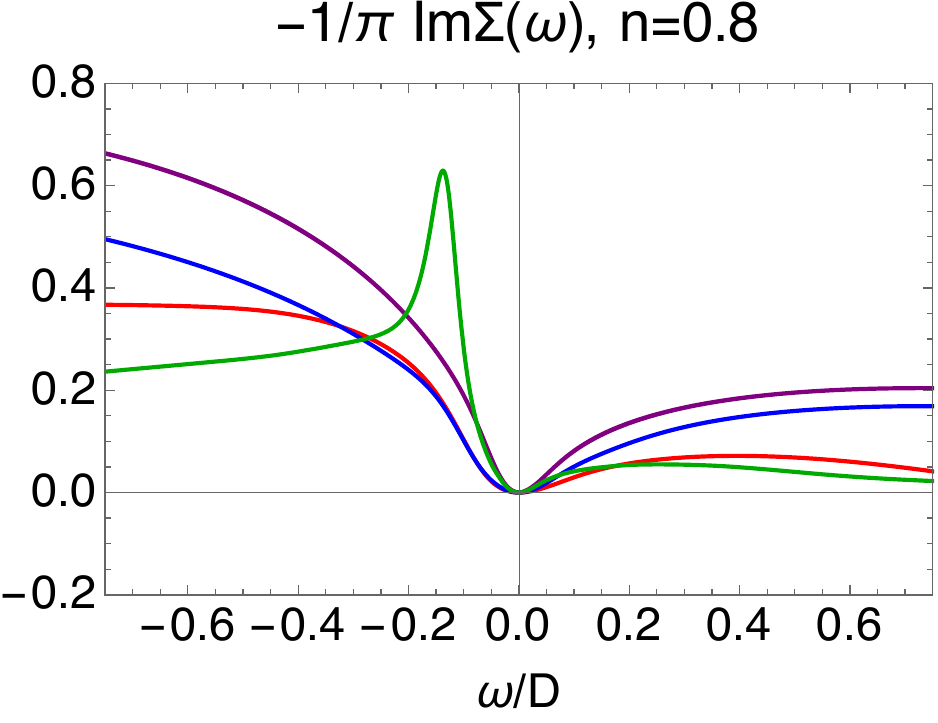}\hfill
\includegraphics[width=.5\textwidth]{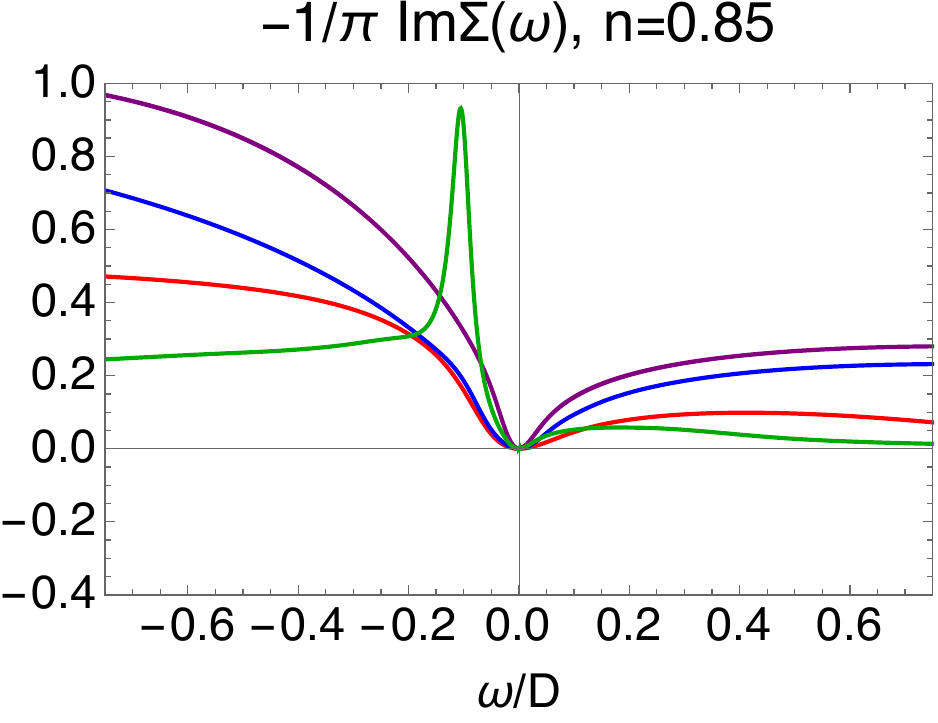}
\caption{\footnotesize The imaginary part of the Dyson self energy for densities 0.7, 0.8 and 0.85, at  T=0.001D. The red plots are second order with the $u_0$ sum rule, the blue are third order with the $u_0$ sum rule, the green are the DMFT results and the purple are second order without the $u_0$ sum rule using the Tukey window scheme.}
\label{Figure4}
\end{figure}

\begin{figure}[htp]
\centering
\includegraphics[width=.33\textwidth]{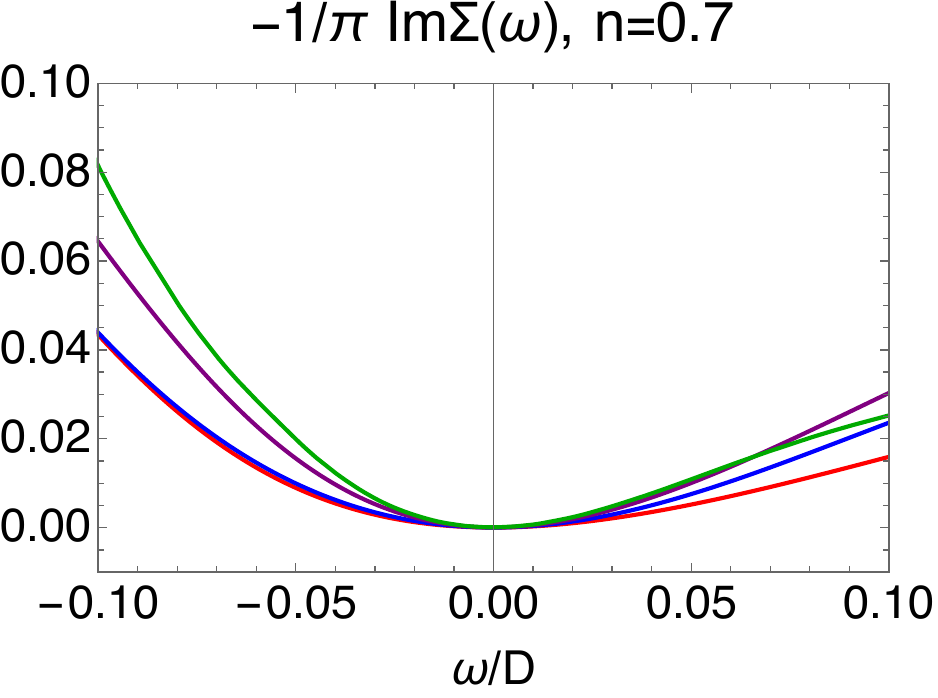}\hfill
\includegraphics[width=.33\textwidth]{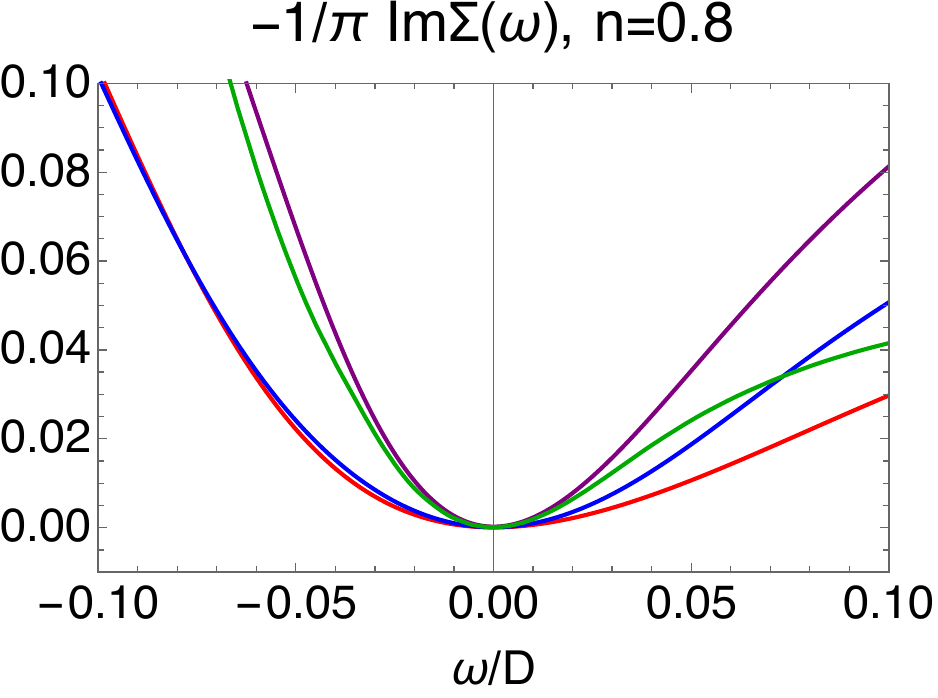}\hfill
\includegraphics[width=.33\textwidth]{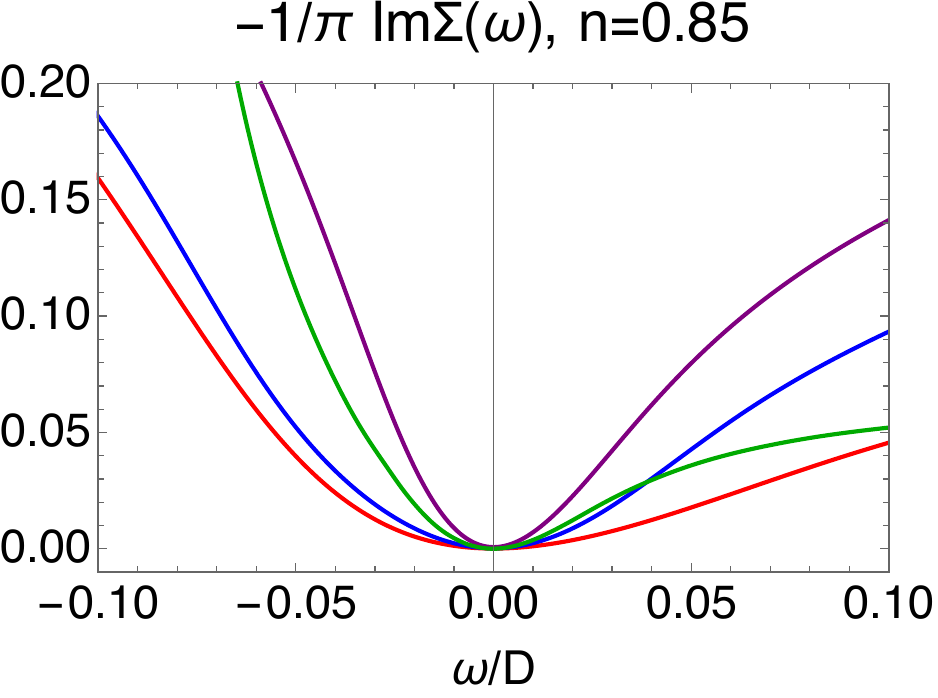}
\caption{\footnotesize Spectral function plots for densities 0.7, 0.8 and 0.85, at  T=0.001D for a smaller frequency range. The red plots are second order with the $u_0$ sum rule, the blue are third order with the $u_0$ sum rule, the green are the DMFT results and the purple are second order without the $u_0$ sum rule using the Tukey window scheme.}
\label{Figure5}
\end{figure}

In \figdisp{Figure4} we display the imaginary part of the Dyson self energy  $\rho_\Sigma (\omega)=-\frac{1}{\pi}\Im\Sigma(\omega)$ at low temperature ($T=0.001$D).
In \figdisp{Figure5} these results are shown over a smaller energy scale $|\omega|\leq 0.1$D  highlighting the lowest lying excitations of the electrons. 
 Our results show a Fermi liquid type  quadratic shape near zero frequency that lines up well with DMFT results. Note that  these plots display spectral asymmetry between particle and hole type excitations, as previously discussed \cite{Asymmetry,ECFL-DMFT-1}. 
 
 In \figdisp{Figure4} we observe a pronounced peak in the DMFT self energy for the somewhat high energy excitations $\omega\sim - 0.2$D. This peak  is missing in all of our ECFL estimates. As a consequence the DMFT electron spectral functions $\rho_\G(k,\omega)$ in \disp{spectral-G} are   more compact in $\omega$ than any of the ECFL estimates on the $\omega<0$ (i.e. occupied) side. 

\begin{figure}[htp]
\centering
\includegraphics[width=.5\textwidth]{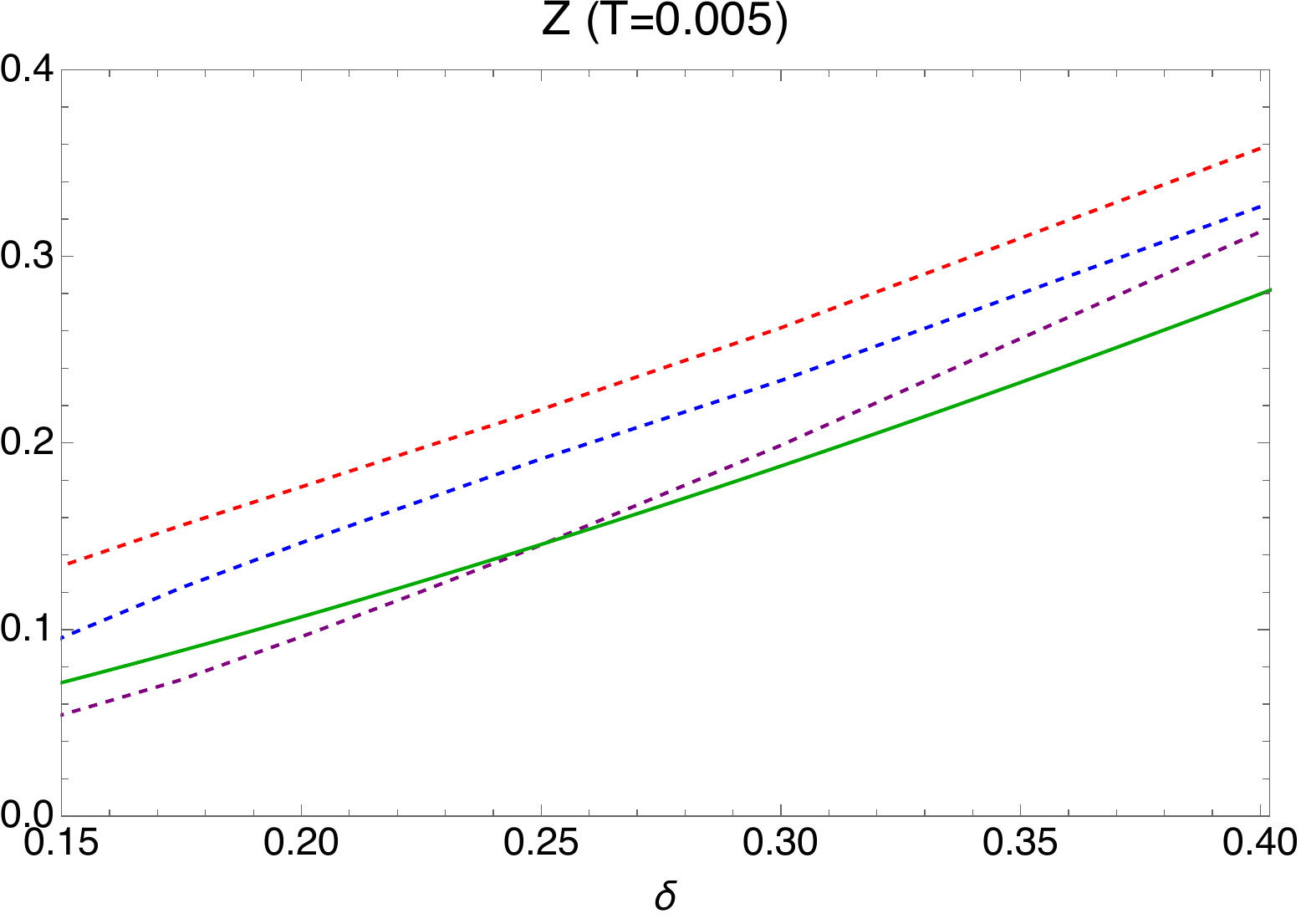}\hfill
\caption{\footnotesize Z for densities 0.7-0.85 ($\delta$ 0.15-0.3).The red plots are second order with the $u_0$ sum rule, the blue are third order with the $u_0$ sum rule, the green are the DMFT results  which fit well to  $Z= \delta^{1.39}$\cite{ECFL-DMFT-1}, and the purple are second order without the $u_0$ sum rule using the Tukey window schemes.}
\label{Figure6}
\end{figure}

The quasiparticle weight $Z$ is obtainable from the self energy as $Z= \{1- \frac{\partial}{\partial \omega} \Re \Sigma(\omega)\}^{-1}$. The strong correlation physics problem usually leads to fragile quasiparticles, i.e. $Z\ll 1$  in the proximity of the Mott-Hubbard insulator at $\delta=0$. The reduction of its magnitude (from unity for the Fermi gas)  is of especial interest, since it is one of the primary  causal agents for the unusual transport and spectral properties in  strongly correlated matter. The calculated $Z$  is  displayed  in \figdisp{Figure6} as a function of the  hole density ($\delta=1-n$) and is seen to be $\ll 1$ as $\delta\leq 0.25$.
 The  Z  from our calculations compares quite well to the DMFT results. As noted earlier  \cite{ECFL-DMFT-1}, the latter  are well fit by $Z\sim\delta^{1.39}$. The third order $u_0$ sum rule results are  closer  to the DMFT results for Z than the second order results { at all densities}, and both of these overestimate the $Z$  for $\delta \leq 0.25$.  In comparison the Tukey window scheme results are closer to the DMFT results but underestimate $Z$ for $\delta \leq 0.25$. { As in the case of the resistivity, the Tukey scheme becomes an overcorrection at $\delta=0.15$.}

%Finally, in Fig. 7, we check to see how well Luttinger's theorem is satisfied in our results. It turns out the Fermii volume is reasonably well preserved, particularly in the 3rd order results.

% \begin{figure}[htbp]
%\centering
%\includegraphics[width=.9\columnwidth]{NCCO.pdf}
 %\caption{\footnotesize  }
% \end{figure}

%\begin{figure}[htp]
%\centering
%\includegraphics[width=.5\textwidth]{DeltaLW.pdf}\hfill
%\caption{\footnotesize Luttinger Ward volume changes compared between 2nd order (red) and 3rd order (blue).}
%\label{Figure6}
%\end{figure}

\section{Concluding Remarks}

%\newpage
%\vspace{.25 in}
The ECFL theory has been developed so far using the $\O(\lambda)$ expressions for the  self energies in the problem, and applied in a variety of situations including $d=0$ i.e. the Anderson Impurity model, the $d=1$ \tJ model, the $d=\infty, U=\infty$ Hubbard model and closest to experiments, the $d=2$ \tJ model. At a formal level we have also established a systematic method for extending the expansion to high order terms, but in view of the additional  technical difficulties presented by them,  these  have not yet been tested. 
This work reports the first results from the third order equations for the ECFL, applied to the case of the $d=\infty$ and $U=\infty$ Hubbard model, where independent DMFT results  are available from the numerical renormalization group. This enables us to quantify the role of the third order terms, and to compare  with the second order results. 

{ The introduction of an exact sum-rule for the \tJ model allows us to bypass the somewhat ad-hoc Tukey window cutoff scheme used in previous ECFL resistivity computations \cite{ECFL-Tukey,ECFL-Dinfty-1,ECFL-Dinfty-2}. In both the case of the second and third order results, the resistivity curve from ECFL agrees in both shape and scale with the one from DMFT, with a quadratic in temperature Fermi-liquid regime, followed by a quasi-linear strange-metal regime. Both ECFL and DMFT predict a monotonic decrease in the quasi-particle weight as one approaches half-filling. In both the case of resistivity and quasi-particle weight, third order ECFL improves upon the second order ECFL at all densities, in comparison to DMFT. The Tukey scheme constitutes a further correction at lower densities, but at higher densities it constitutes and over-correction, overshooting the DMFT results. Finally, both ECFL and DMFT find the quadratic quasi-particle minimum in the Dyson self-energy at low frequencies, while DMFT has a higher (negative) frequency peak, which is absent from the low-order ECFL results. It is encouraging that in going from second to third order in the ECFL computation we obtain better agreement with DMFT.

 }

\vspace{.25 in}

\section{Acknowledgements} 

We are  grateful to Professor Rok \v{Z}itko for permitting us to use the results of  his dynamical mean field theory calculations of  the $d=\infty$ and $U=\infty$  data for comparison with our results. The work at UCSC was supported by the US Department of Energy (DOE), Office of Science, Basic Energy Sciences (BES), under Award No. DE-FG02-06ER46319. The computation was done on the comet in XSEDE\cite{xsede} (TG-DMR170044) supported by National Science Foundation grant number ACI-1053575.

\appendix

\section{Appendix: The second sum rule \label{Sumrule}}

We give a brief derivation of the sum-rule valid in infinite dimensions
\beq
\sum_k \int d\omega \rho_{\G}(k,\omega) f(\omega) (\omega+ \chem- \varepsilon_k) =0. \label{second-sumrule}
\eeq

We start with the Hamiltonian (energy minus $\chem N$) written in terms of the Hubbard operators\cite{ECFL-0,ECFL-Main}
\beq 
H=-\sum_{i,j,\sigma}t_{ij}X_i^{\sigma 0} X_j^{0\sigma}-\chem \sum_{i,\sigma} X_i^{\sigma \sigma}+\frac{1}{2}\sum_{i,j}J_{ij}(\vec{S}_i\cdot\vec{S}_j-\frac{1}{4}n_i n_j)\eeq
We rewrite this in the form
\beq=-\sum_{i,j,\sigma}t_{ij}X_i^{\sigma 0} X_j^{0\sigma}-\chem \sum_{i,\sigma} X_i^{\sigma \sigma}+ V_{ex} \eeq
where the exchange energy is 
\beq
V_{ex}&=&  - \frac{1}{4} \sum_{ij \si \si'} \si \si' J_{ij} X_i^{\si \si'} X_j^{\sib \sib'} \nn\\
&=&\frac{1}{4}\sum_{ij,\sigma}J_{ij}(X_i^{\sigma \overbar{\sigma}} X_j^{\overbar{\sigma}\sigma}-X_i^{\sigma \sigma} X_j^{\overbar{\sigma}\overbar{\sigma}})
\eeq

and define the electron Green's function
\beq
%\G_{\si_i  \si_f}(i \tau_i,f \tau_f)= - \langle\langle X_{i}^{0 \si_i}(\tau_i) X_{f}^{\si_f 0}(\tau_f) \rangle \rangle = - \frac{1}{Z_G} \Tr e^{- \beta H} T_\tau\left( X_{i}^{0 \si_i}(\tau_i) X_{f}^{\si_f 0}(\tau_f) \right)
\G_{\si_i  \si_f}(i \tau_i,f \tau_f)= - \frac{1}{Z_G} \Tr e^{- \beta H} T_\tau\left( X_{i}^{0 \si_i}(\tau_i) X_{f}^{\si_f 0}(\tau_f) \right)
\eeq
where  $Z_G= \Tr e^{- \beta H}$.
Taking the time derivative with respect to $\tau_i$ and then setting $\tau_f=\tau_i+0^+$,  $\si_i=\si_f=\si$ and the sites  $i=f$   we get
\beq
\partial_{\tau_i}\G_{\si \si}(i,i^+) = \chem \langle X_i^{\si \si}\rangle + \sum_{j} t_{ij} \langle X_i^{\si 0} X_{j}^{0 \si} \rangle + \frac{1}{2} \sum_{j \si'} J_{ij} \si \si' \langle X_i^{\si \si'} X_j^{\sib \sib'} \rangle
\eeq
where we dropped a term containing $\delta(\tau_i-\tau_f)$ (since we are considering the limit 
$\tau_f=\tau_i+0^+$). Summing over $\si$, denoting  $\tau= \tau_i-\tau_f$,  and summing over  site index $i$ (replaced by a $\vec{k}$ sum),  we get
\beq
2 \partial_{\tau}\sum_{\vec{k}} \G(\vec{k},\tau) \bigg\vert_{\tau\to 0^-}= \chem N - \langle T\rangle - 2 \langle V_{ex} \rangle.\label{EOM-2}
\eeq
It is convenient to introduce the general formula for the Greens function in terms of the spectral function in time domain
\beq
\G(\vec{k},\tau) = \int \, d\nu \, \rho_{\G}(\vec{k},\nu) \, e^{- \nu \tau} \left( f(\nu) \Theta(-\tau) - \bar{f}(\nu) \Theta(\tau) \right),
\eeq
where $f$ is the fermi function and $\bar{f}=1-f$, and $\Theta(\tau)$ is the Heaviside theta function. Substituting  into \disp{EOM-2} and transposing terms we get
\beq
-2 \sum_{\vec{k}} \int \, d\nu \, \nu \,  \rho_{\G}(\vec{k},\nu) -  \chem N + \langle T\rangle= - 2 \langle V_{ex} \rangle
\eeq
Using $ 2 \sum_{\vec{k}} \int \, d\nu \,  \,  \rho_{\G}(\vec{k},\nu)=N$ and $  2 \sum_{\vec{k}} \int \, d\nu \, \varepsilon_k \,  \rho_{\G}(\vec{k},\nu)=\langle T\rangle$, we get

\beq
 \sum_k \int d\omega \rho_{\G}(k,\omega) f(\omega) (\omega+ \chem- \varepsilon_k) =  \langle V_{ex} \rangle. 
\eeq
In $d=\infty$ we set $J=0$ and hence $V_{ex}$ vanishes and we get the sum-rule \disp{second-sumrule}. 

In lower dimensions we are obliged to use some suitable  approximation to estimate $\langle V_{ex}\rangle$. In the physically important case relevant to cuprates of $d=2$ on a square lattice (with 4 neighbors), we 
may use a Hartree type approximation
\beq
\langle V_{ex} \rangle =   - \frac{1}{4} \sum_{ij \si \si'} \si \si' J_{ij} \langle X_i^{\si \si'} X_j^{\sib \sib'}\rangle \sim  - \frac{1}{4} \sum_{ij \si }  J_{ij} \langle X_i^{\si \si} \rangle \langle X_j^{\sib \sib} \rangle 
= -
\frac{n^2}{8}  N_s Z_c J,  \eeq
where $J$ is the  nearest neighbor exchange energy and $Z_c$ is the number of nearest neighbors in the lattice.

\section{Appendix: Program Notes \label{Code}}

Our program at both second and third order uses a rootfinder with two equations and two variables to solve for $\chem'$and $u_0$.
We use \disp{Second-sumrule,First-sumrule} as mentioned in the text.
We noted that  the third order program is significantly more stable with this choice of sum-rules. 

 It is generally true that, whichever two sumrules are chosen, the third will be approximately satisfied. Since the $n_G$ rule and new $u_0$ sumrule are used, $n_G$ is exactly equal to $n$, while $n_g$ is only approximately equal to $n$. As mentioned previously, the $n_g$ value generally ends up $10$ to $15$\% higher than $n$. When used in \disp{Eq-28}, the different value of $n_g$ can cause noise under iteration, resulting in a failure to converge. This effect is particularly pronounced for the $O(\lambda^3)$ program. So we approximate $n_g$ with $n$ in our chemical potential (\disp{chemical}), which gives very similar results in all well behaved cases we compared. It should be noted that multiplying $n$ by a constant to bring it closer to $n_g$ also causes failure to converge at third order; for best results the $n$ approximation should be used.

Here we would also like to outline the parameters under which our programs are well behaved. The $O(\lambda^2)$ program converges with relative ease for a wide range of temperatures and densities. We tested densities around 0.5-0.9 and temperatures from 0.001 to 0.2 with good results. The $O(\lambda^2)$ program also functions well with the $n_g$ sumrule substituted for the $n_G$ sumrule.

The third order program is generally more unstable than second order. It converges comfortably for densities 0.7-0.85 over our full temperature range, 0.001-0.02. Beyond those densities the program rapidly becomes more difficult to run. For lower densities it is possible to push the program to converge a little below 0.6. For higher densities in particular the third order program consistently has significant difficultly converging. We recommend this technique not be extended beyond optimal density (0.85).


\begin{thebibliography}{99}
\bibitem{FN-1} These follow from Eq.~(7,8,9) of \cite{ECFL-Tukey}, where we  replaced $a_G$ by $ 1-\lambda n/2$ using the number sumrule \disp{First-sumrule}  for the physical particles.
\bibitem{comment-1} This sum-rule  is a consequence of the $\frac{1}{\ \omega}$ fall off of $\GH$ (see \disp{Eq-g}, and $\lim_{\omega \to \infty} \Phi(i \omega) \to 0$).
\end{thebibliography}
 \end{document}